\documentclass[a4paper,serif,12pt]{article}
\usepackage{geometry}
\usepackage{makeidx}
\usepackage{graphicx}
\usepackage[dvipsnames,svgnames,table]{xcolor}
\usepackage{epstopdf}
\usepackage{epsfig}
\usepackage[font=small]{caption}
\usepackage{textcomp}
\usepackage{hyperref}
\usepackage{amsmath}
\usepackage{amssymb}
\usepackage{multirow}
\usepackage{multicol}
\usepackage{booktabs}
\usepackage{lastpage}
\usepackage{hyperref} 
\usepackage{enumerate}
\usepackage{threeparttable}
\usepackage{float}
\usepackage{array}
\usepackage{cite}
\makeatletter
\renewcommand\section{\@startsection{section}{1}{\z@}%
                                       {-2.5ex \@plus -1ex \@minus -.2ex}%
                                       {2.3ex \@plus.2ex}%
                                       {\normalfont\large\bfseries}}
\renewcommand\subsection{\@startsection{subsection}{1}{\z@}%
                                       {-2.5ex \@plus -1ex \@minus -.2ex}%
                                       {2.3ex \@plus.2ex}%
                                       {\small\bfseries}}
\makeatother
\begin{document}
\begin{center}
\large{\textbf{Neutronic Analysis of a Nuclear-Chicago NH3 Neutron Howitzer}}
\end{center}
\bigskip
\begin{center}
\noindent\textbf{Ahmet Ilker Topuz, Iskender Atilla Reyhancan }\\
\medskip
Istanbul Technical University, Energy Institute, 34469 Istanbul, Turkey\\
E-mail address:  topuz15@itu.edu.tr, aitopuz@gmail.com\\
Tel:  +90 534 020 70 00\\
\begin{abstract}
Neutron howitzers are the irradiation instruments, generally used for the chemical analysis in the research laboratories, where the neutron sources are encapsulated within a moderating medium and the target materials are irradiated at the channels inserted through the corresponding neutron sources. In this study, we employ Monte Carlo simulations by using the MCNP5 code in order to perform the neutronic analysis of a Nuclear-Chicago NH3 neutron howitzer with a 5-Ci ${}^{239}$Pu-Be neutron source. This neutron howitzer is a cylindrical drum that is covered by aluminum at the level of outermost layer, and the cylindrical body is filled with paraffin for the neutron moderation. The drum consists of two vertical and two horizontal irradiation channels and a vertical source channel in which the 5-Ci ${}^{239}$Pu-Be neutron source is located and moved upward or downward by the aid of a plexiglass rod. We focus our study on the two horizontal irradiation channels, separated by an angle of 120$^\circ$, which are infilled with a cylindrical plexiglass bar. We construct our geometry by respecting the dimensional properties of the NH3 howitzer components and we define a biased cylindrical source of 1.02 in.$\times$4.425 in. by taking into account the neutron energy spectrum and source strength of the 5-Ci ${}^{239}$Pu-Be neutron source. We determine the neutron flux within the plexiglass bar for various energy bins in the energy range between 0 and 10.5 MeV, and also, we investigate the axial variation of the neutron flux along the plexiglass region in the horizontal irradiation channels. Our simulation results show that a significant number of thermal neutrons as well as a non-negligible population of fast neutrons propagate through the plexiglass bar, thereby demonstrating the opportunity to profit from a wide range of energy spectrum for the neutron activation analysis.
\end{abstract}
\end{center}
\section{Introduction}
Elemental analysis of the raw or processed materials is a fundamental task that is repeatedly performed during the life cycle from the extraction to the disposal or the recycling/recovery when possible. While this chemical analysis might be implemented through the medium of several different methods, nuclear techniques have also been introduced to determine the elemental composition. Among the existing nuclear practices is the neutron activation analysis that is founded on the bombardment of the target material by a beam of incident neutrons~\cite{acti1,acti2,acti3}.

The neutron activation analysis is usually carried out and deliberately labeled for a particular energy group (e.g. thermal, epithermal, or fast), and the activity of the irradiated material linearly depends on the neutron flux of this particular energy group. Thus, the energy spectrum of the neutron beam and the magnitudes of the neutron flux within the specific energy groups are of technical importance. Accordingly, research reactors are frequently employed owing to the high flux values~\cite{reactorna1,reactorna2,reactorna3}, whereas neutrons generators are generally preferred for the mid-scale irradiation campaigns~\cite{ngen1,ngen2,ngen3}.

Neutron howitzer is a sort of a neutron generator providing the neutron emissions based on the isotopic neutron sources~\cite{PuBeStr1,PuBeStr2,isotopic}, and Nuclear-Chicago NH3 is acknowledged to be a notable neutron howitzer that was offered as a commercial product for the chemical analysis in the small scale laboratories~\cite{Howit1,Howit2,Howit3}. The isotopic neutron source that is commonly used in the NH3 howitzer is $^{239}$Pu-Be source, and the generation of neutrons is sustained by virtue of ($\alpha$,n) reaction in a cylindrical drum filled with ordinary paraffin~\cite{chic1,chic2,chic3}.  Within the body of the NH3 howitzer, two vertical irradiation ports and two horizontal irradiation ports are available; however, the horizontal irradiation channels (i.e. ports) are more frequently operated when the channel diameter and the distance from the neutron source are taken into consideration.

In the current study, we employ Monte Carlo simulations (MCS) by using the MCNP5 code to explore the neutronic aspects of the horizontal beam port filled with a plexiglass bar inside our Nuclear-Chicago NH3 howitzer with a 5-Ci ${}^{239}$Pu-Be neutron source~\cite{mcnp}. First, we construct a 3D geometry by taking the components of the NH3 howitzer into account, and then, we entirely focus on the plexiglass bar of the horizontal beam port by using a strategy based on geometric splitting with Russian roulette that accents the neutron transfer region consisting of the vertical source channel, the paraffin moderator, and the horizontal irradiation channel. Thereafter, we distribute a number of point detectors~\cite{briesmeister} along the axial distance within one of the horizontal irradiation channels. Finally, we obtain the axial variation of the neutron flux for a number of energy bins as well as the thermal, epithermal, and fast energy intervals besides the collided neutron flux and the uncollided neutron flux.
\section{Nuclear-Chicago NH3 howitzer with 5-Ci ${}^{239}$Pu-Be}
\begin{figure}[H]
\begin{center}
\setlength{\belowcaptionskip}{-4ex} 
\includegraphics[width=7.5cm]{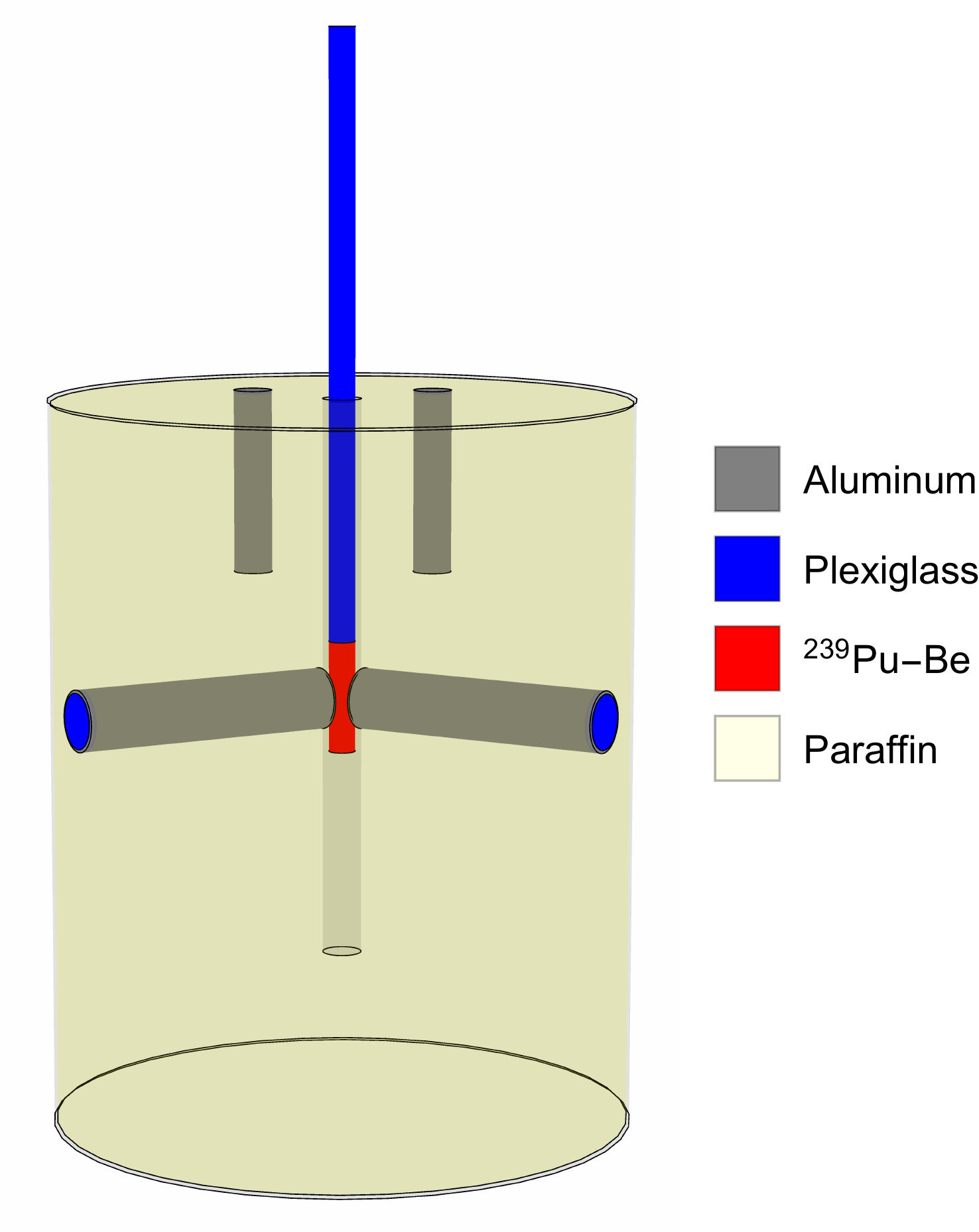}
\caption{Layout of the Nuclear-Chicago NH3 howitzer with the 5-Ci ${}^{239}$Pu-Be source.}
\label{Howitzer}
\end{center}
\end{figure}
The Nuclear-Chicago NH3 howitzer is an aluminum cylindrical drum that comprises two types of irradiation channels; namely, the horizontal irradiation channel and the vertical irradiation channel. The separation angle between the horizontal irradiation channels is 120$^\circ$, and the source channel along with the two vertical irradiation channels makes an isosceles right triangle. The layout of the NH3 howitzer is depicted in Fig.~\ref{Howitzer}.
\begin{table} [H]
\begin{footnotesize}
\begin{center}
\begin{threeparttable}
\captionsetup{skip=0pt}
\caption{Structural properties of the Nuclear-Chicago NH3 components.}
\begin{tabular}{*5c}
\toprule
Components&Material&Geometry&Radius (cm)$\times$Height (cm)&Thickness (cm)\\
\midrule
Drum& Aluminum&Closed cylinder&28.8\tnote{a} $\times$72 & 0.3\\
Moderator& Paraffin&Closed cylinder&28.5$\times$71.4 &57\tnote{b} \\
Source channel&Aluminum&Open cylinder&1.9\tnote{a} $\times$55.3 &0.3\\
Source rod&Plexiglass&Closed cylinder&1.3$\times$57.8&2.6\tnote{b}\\
Vertical channel&Aluminum&Open cylinder&1.9\tnote{a} $\times$18.3&0.3\\
Horizontal channel&Aluminum&Open cylinder &3\tnote{a} $\times$27.2&0.3\\
Horizontal channel bar&Plexiglass&Closed cylinder& 2.7$\times$26.9&5.4\tnote{b}\\
\bottomrule
\end{tabular}
\label{geodim}
\nointerlineskip
\begin{tablenotes}
\item[a] Outer radius
\item[b] Diameter
\end{tablenotes}
\end{threeparttable}
\end{center}
\end{footnotesize}
\end{table}
\vspace*{-\baselineskip}
The cylindrical drum of the NH3 howitzer is loaded with ordinary paraffin (15.7 wt.$\%$H+84.3 wt.$\%$C) that serves as the moderating medium for the purpose of the neutron thermalisation. While the vertical beam ports are empty, plexiglass (8 wt.$\%$H+32 wt.$\%$O+60 wt.$\%$C) bars are inserted into the horizontal beam ports. The 5-Ci ${}^{239}$Pu-Be neutron source, which is hold by a plexiglass rod, is accommodated by the source channel. The structural properties of the NH3 components are abstracted in Table~\ref{geodim}. The compendium of material composition data from the PNNL is used for the production of the material card~\cite{compendium}.

\begin{figure}[H]
\begin{center}
\setlength{\belowcaptionskip}{-4ex} 
\includegraphics[width=13cm]{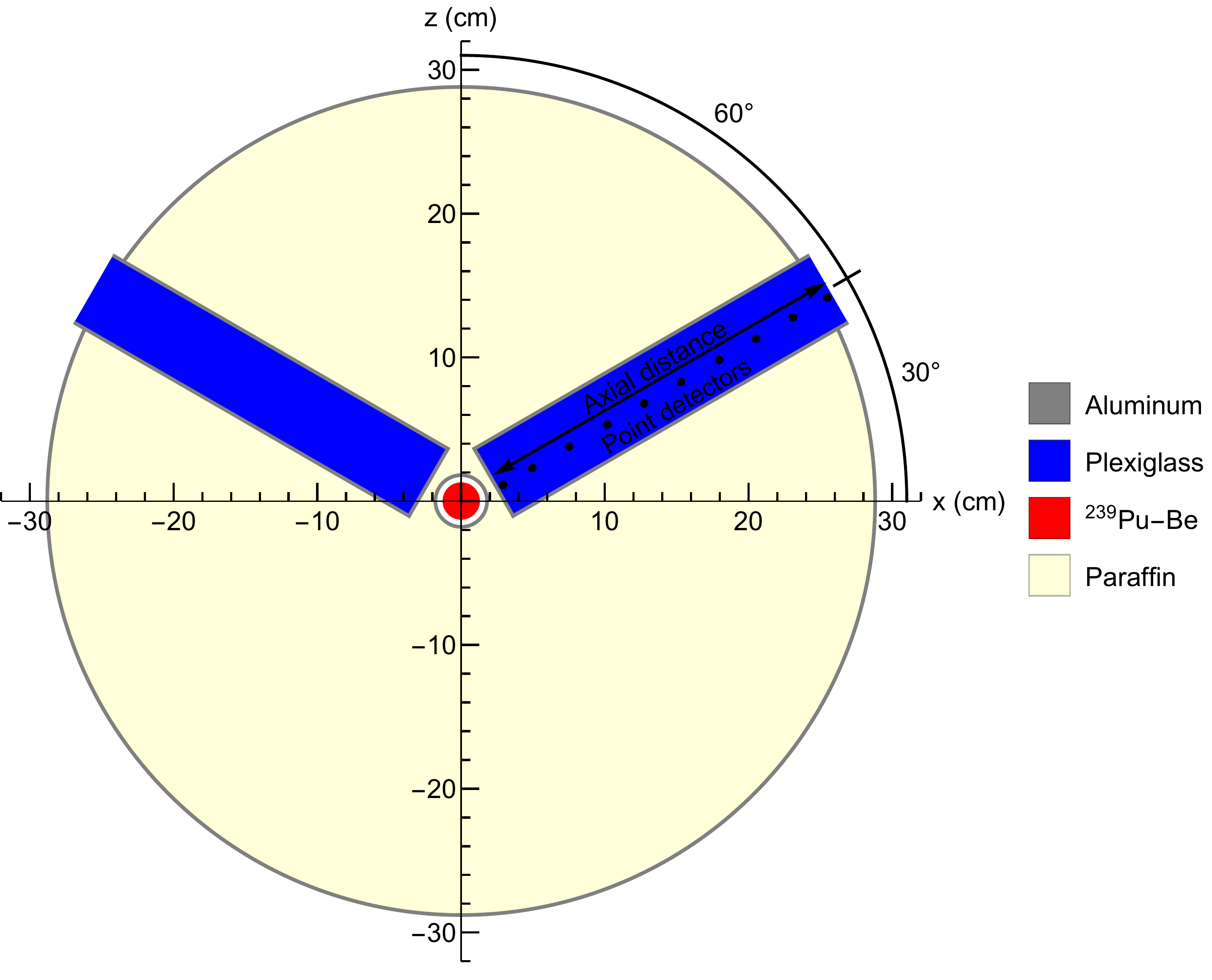}
\caption{2D cross-sectional scheme of the Nuclear-Chicago NH3 howitzer when y=42.5 cm.}
\label{Veraxe}
\end{center}
\end{figure}

In the current study, the center of the 5-Ci ${}^{239}$Pu-Be neutron source is placed aligned with the horizontal beam port, which is equivalent to a height of 42.5 cm from the bottom of the Nuclear-Chicago NH3 howitzer. The size of the  5-Ci ${}^{239}$Pu-Be neutron source is 1.02 in.$\times$4.425 in.~\cite{chic2}. The neutron strength, which is utilised for the normalization of flux values, equals 1.6$\times10^{6}$ $\rm s^{-1}Ci^{-1}$~\cite{PuBeStr1,PuBeStr2}. The neutron source is biased by applying the radial and axial sampling weights~\cite{Shultis}. The energy spectrum corresponds to the interval between 0 and 10.5 MeV, where the mean energy ($E_{\rm mean}$) is equal to 4.24 MeV~\cite{PuBespec}. The source features are listed in Table~\ref{sourceprop}.  
\begin{table} [H]
\begin{footnotesize}
\begin{center}
\begin{threeparttable}
\captionsetup{skip=0pt}
\caption{Features of the 5-Ci ${}^{239}$Pu-Be in the current study.}
\begin{tabular*}{\columnwidth}{@{\extracolsep{\fill}}*5c}
\toprule
Source&Geometry&Activity (Ci)&Strength ($\rm s^{-1}Ci^{-1}$)&Diameter (in.)$\times$Height (in.)\\
\midrule
${}^{239}$Pu-Be&Closed cylinder&5&1.6$\times10^{6}$&1.02$\times$4.425\\
\midrule
\end{tabular*}
\begin{tabular*}{\columnwidth}{@{\extracolsep{\fill}}*5c}
\midrule
Radial weighting&Axial weighting&Spectrum (MeV)&Major peaks (MeV)&$E_{\rm mean}$ (MeV)\\
\midrule
1\tnote{a} & 0\tnote{a} & [0,10.5] & 3.1,4.5& 4.24\\
\bottomrule
\end{tabular*}
\label{sourceprop}
\nointerlineskip
\begin{tablenotes}
\item[a] Power law
\end{tablenotes}
\end{threeparttable}
\end{center}
\end{footnotesize}
\end{table}
\vspace*{-\baselineskip}
In our MCS, we concentrate on the horizontal beam port of the NH3 howitzer that is illustrated in Fig.~\ref{Veraxe}. The length of the plexiglass bar in the horizontal irradiation channel is 26.9 cm, and the distance between the source channel and the horizontal beam port is 0.5 cm. Point detectors, which also allow the calculation of the collided neutron flux and the uncollided neutron flux, are distributed along the plexiglass bar~\cite{briesmeister}, and the radius of these point detectors is 0.2 cm~\cite{Shultis}. The geometric splitting with Russian roulette is employed to reduce the uncertainties in the targeted region. The number of histories used in all the simulation cases is $3\times10^{6}$.
\section{Neutronic aspects of the horizontal beam port}
\begin{figure}[H]
\begin{center}
\setlength{\belowcaptionskip}{-4ex} 
\includegraphics[width=12cm]{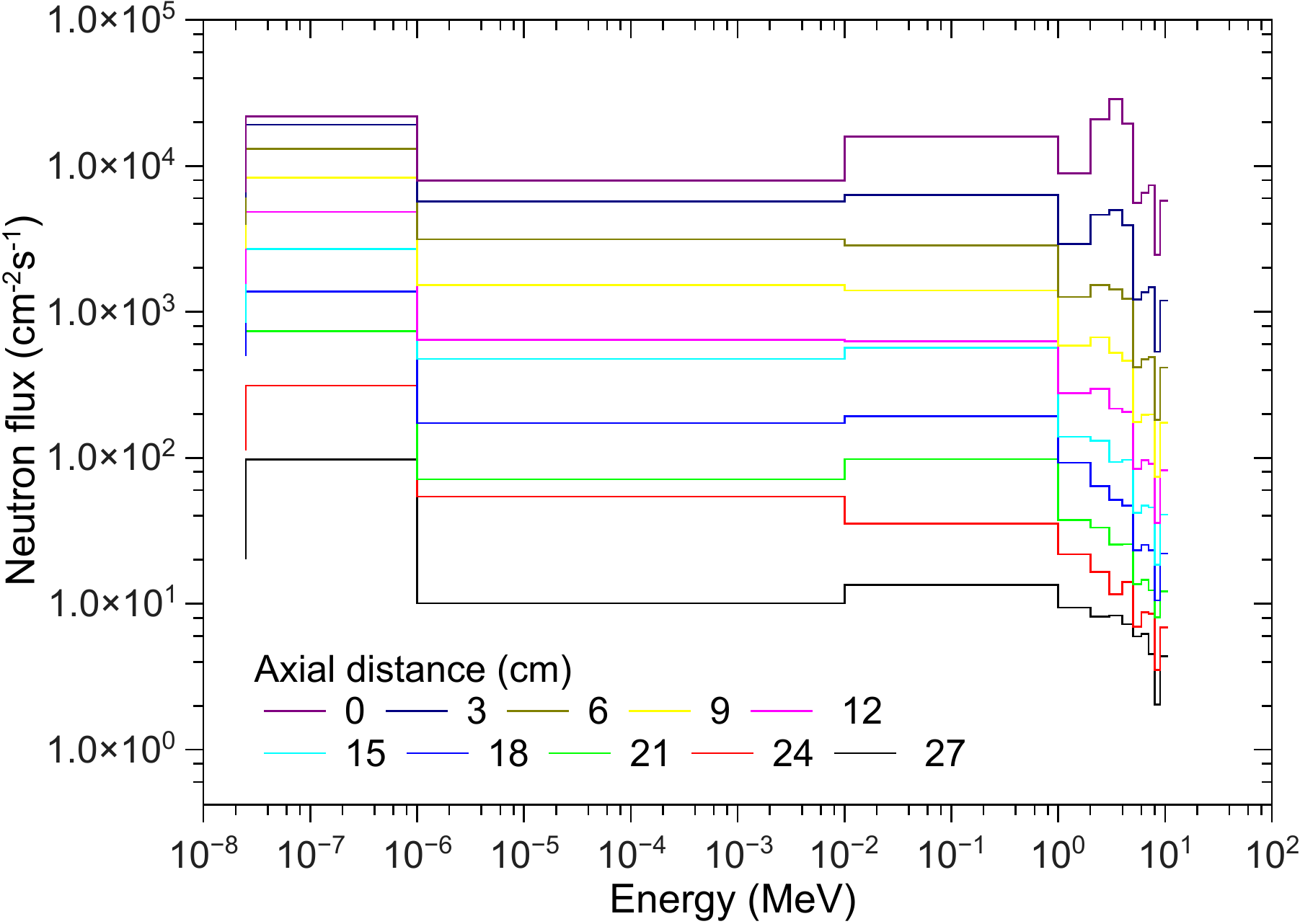}
\caption{Variation of neutron flux-energy bin by using ten point detectors.}
\label{horstepbins}
\end{center}
\end{figure}
\begin{figure}[H]
\begin{center}
\setlength{\belowcaptionskip}{-4ex} 
\includegraphics[width=12cm]{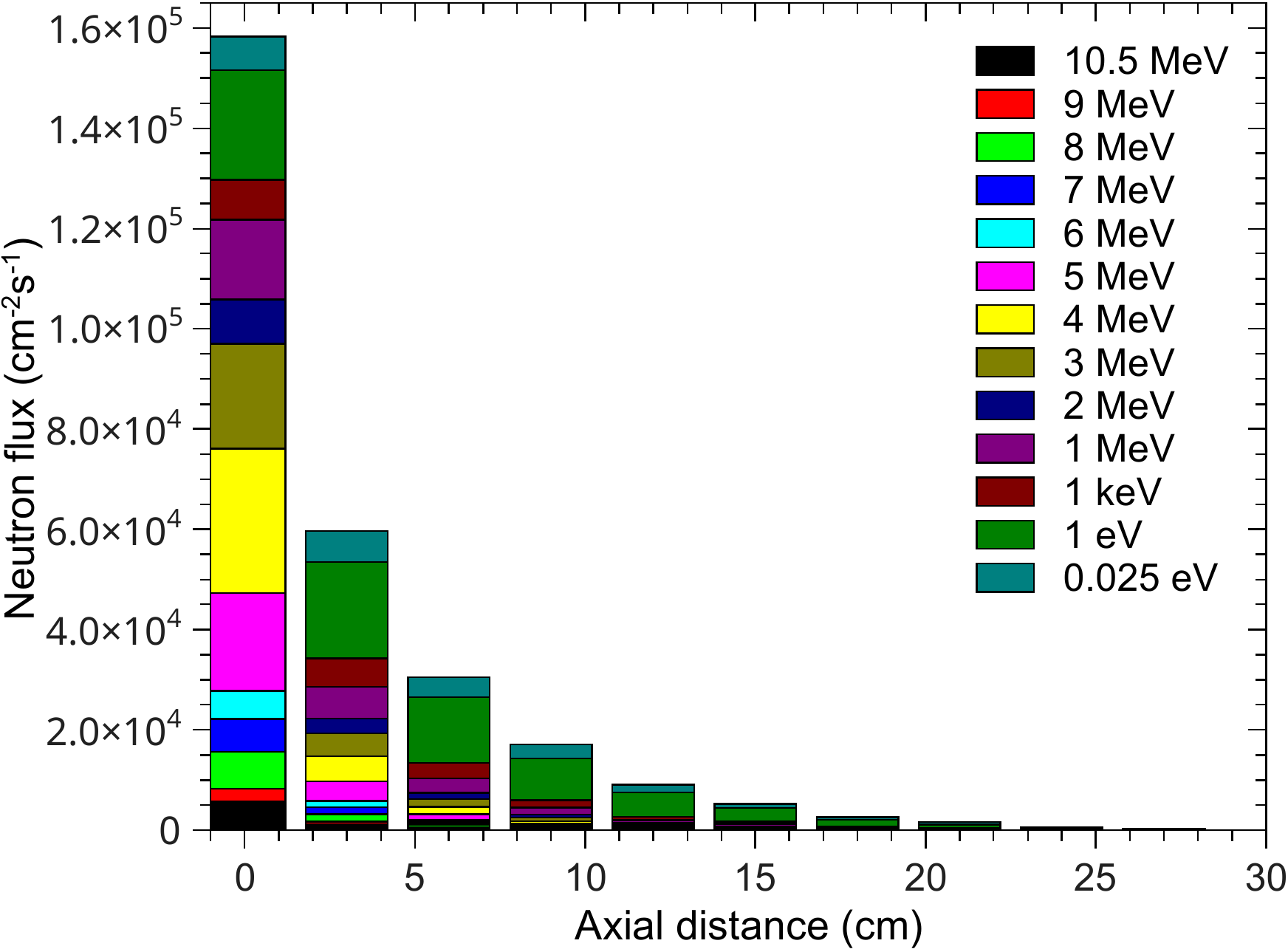}
\caption{Axial variation of the neutron flux accumulated over the defined energy bins.}
\label{stack}
\end{center}
\end{figure}
\begin{figure}[H]
\begin{center}
\setlength{\belowcaptionskip}{-4ex} 
\includegraphics[width=16cm]{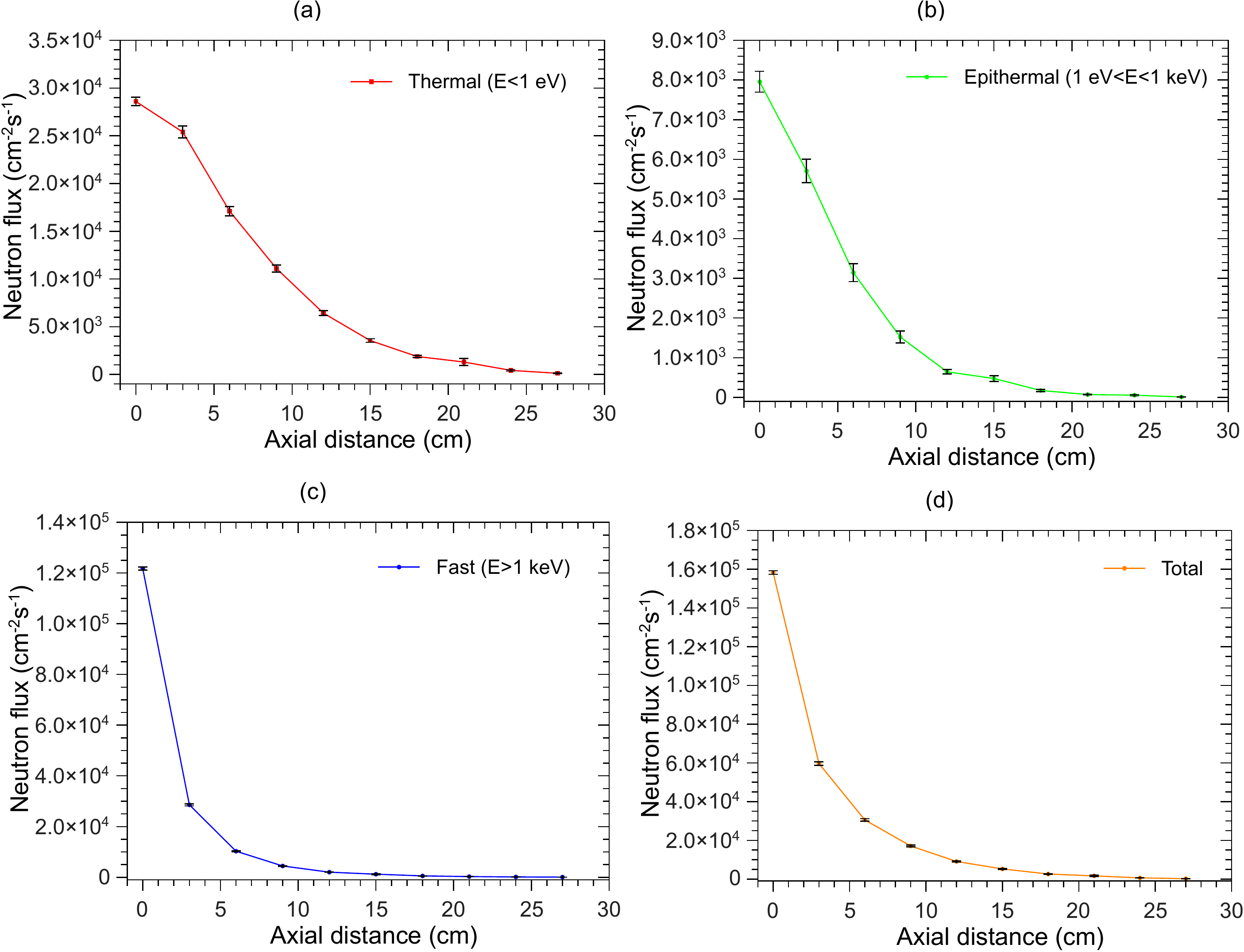}
\caption{Axial variation of the neutron flux for a coarser energy division: (a) thermal, (b) epithermal, (c) fast, and (d) total.}
\label{thermepifastot}
\end{center}
\end{figure}
\begin{figure}[H]
\begin{center}
\setlength{\belowcaptionskip}{-4ex} 
\includegraphics[width=12cm]{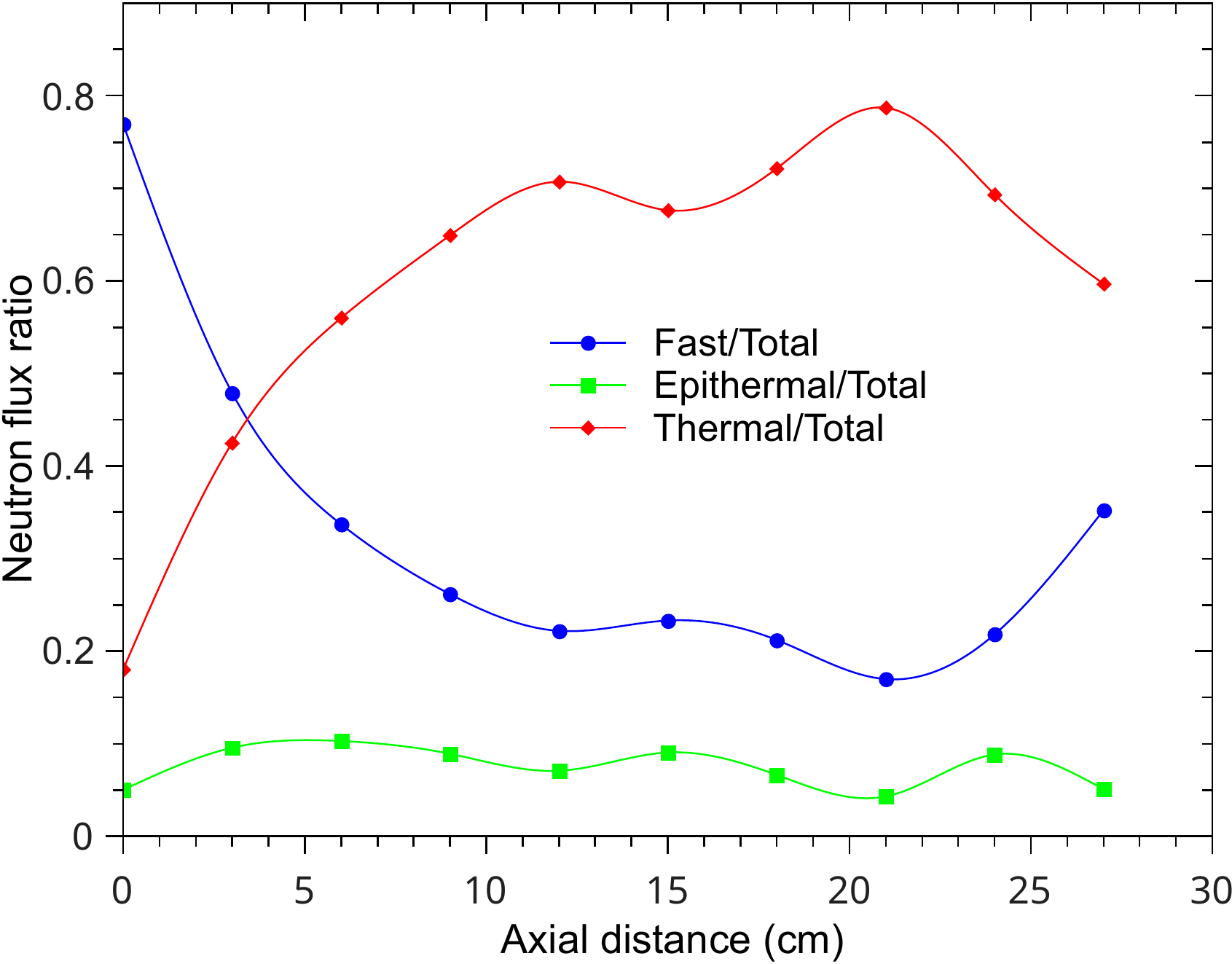}
\caption{Axial variation (spline) of the thermal/total, epithermal/total, and fast/total neutron flux ratios.}
\label{ratio}
\end{center}
\end{figure}
\begin{figure}[H]
\begin{center}
\setlength{\belowcaptionskip}{-4ex} 
\includegraphics[width=12cm]{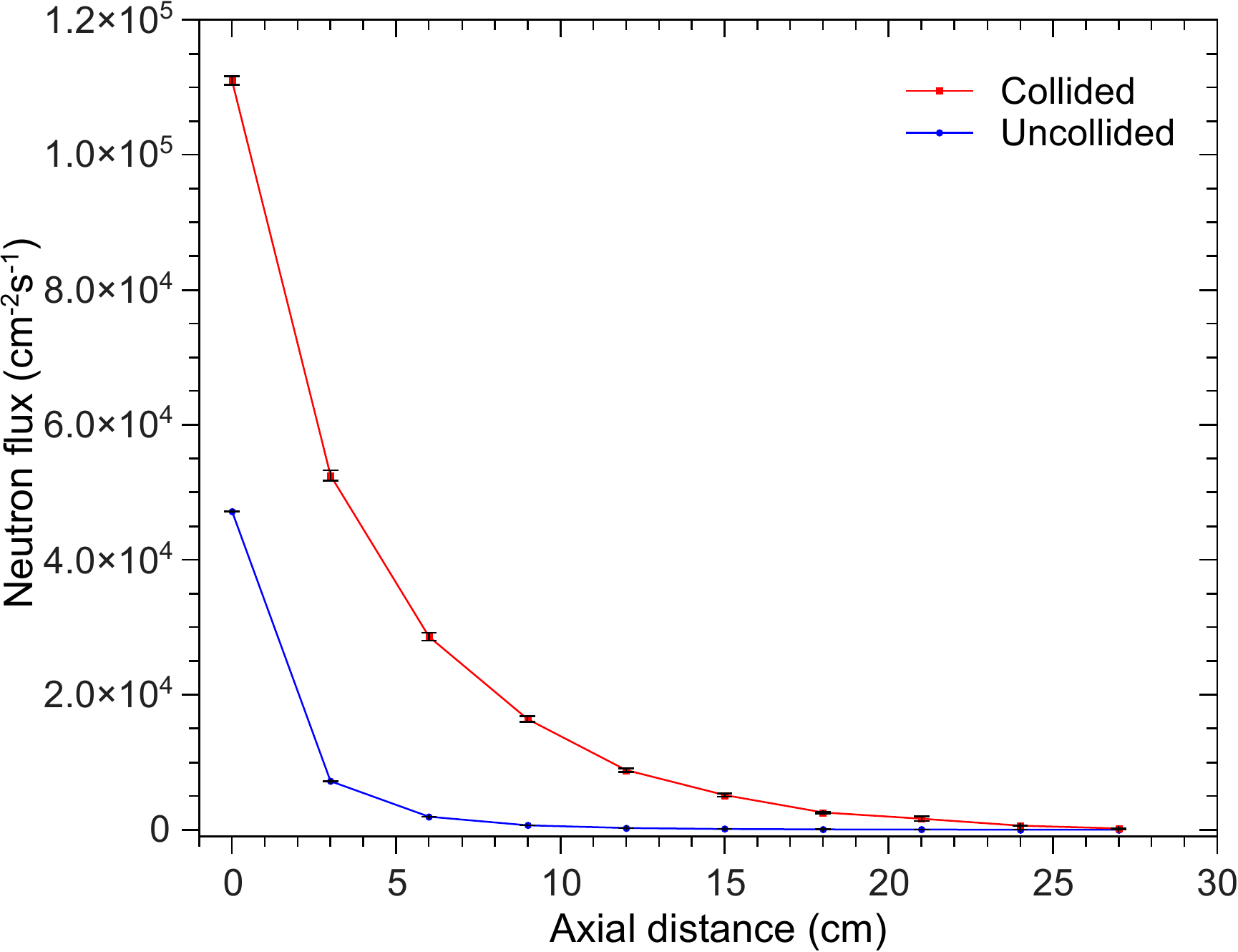}
\caption{Axial variation of the collided and uncollided neutron flux.}
\label{coluncol}
\end{center}
\end{figure}
\begin{figure}[H]
\begin{center}
\setlength{\belowcaptionskip}{-4ex} 
\includegraphics[width=12cm]{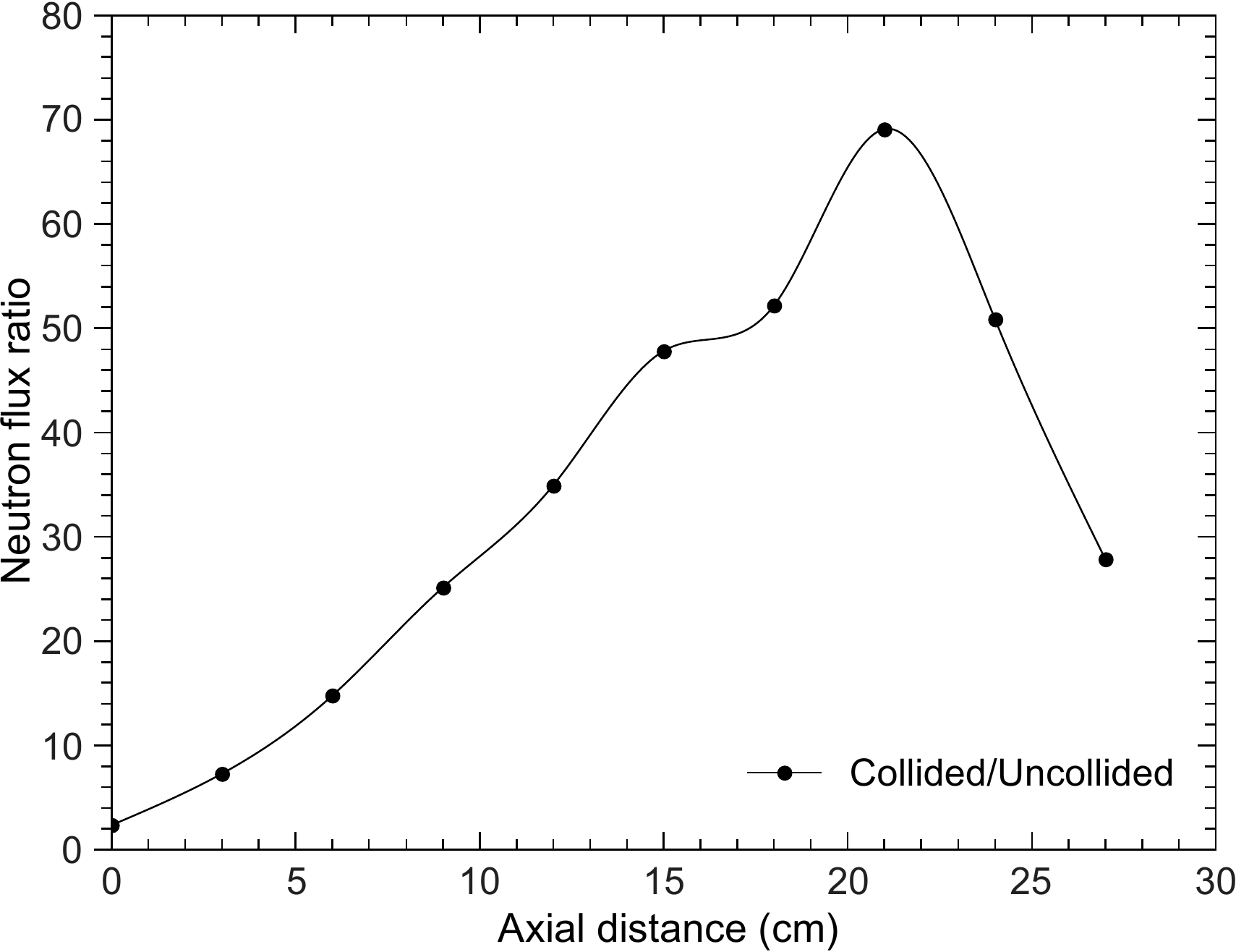}
\caption{Axial variation (spline) of the collided/uncollided neutron flux ratio.}
\label{coluncolratio}
\end{center}
\end{figure}

Throughout the present study, by using the MCS, we investigate the neutronic aspects of the horizontal beam port depicted in Fig.~\ref{Veraxe}. First, we check the variation of the neutron flux along the axial distance in the plexiglass bar by employing ten point detectors for a list of energy bins where the elements of this list gradually increase from 0.025 eV to 10.5 MeV. The change of the neutron flux with respect to these defined energy bins is shown in Fig.~\ref{horstepbins}, and we remark that the fast energy regime is dominant over the first spatial position; however, this characteristic is replaced by the elevation of the thermal regime at most of the tracking points.
\begin{figure}[H]
\begin{center}
\setlength{\belowcaptionskip}{-4ex} 
\includegraphics[width=12cm]{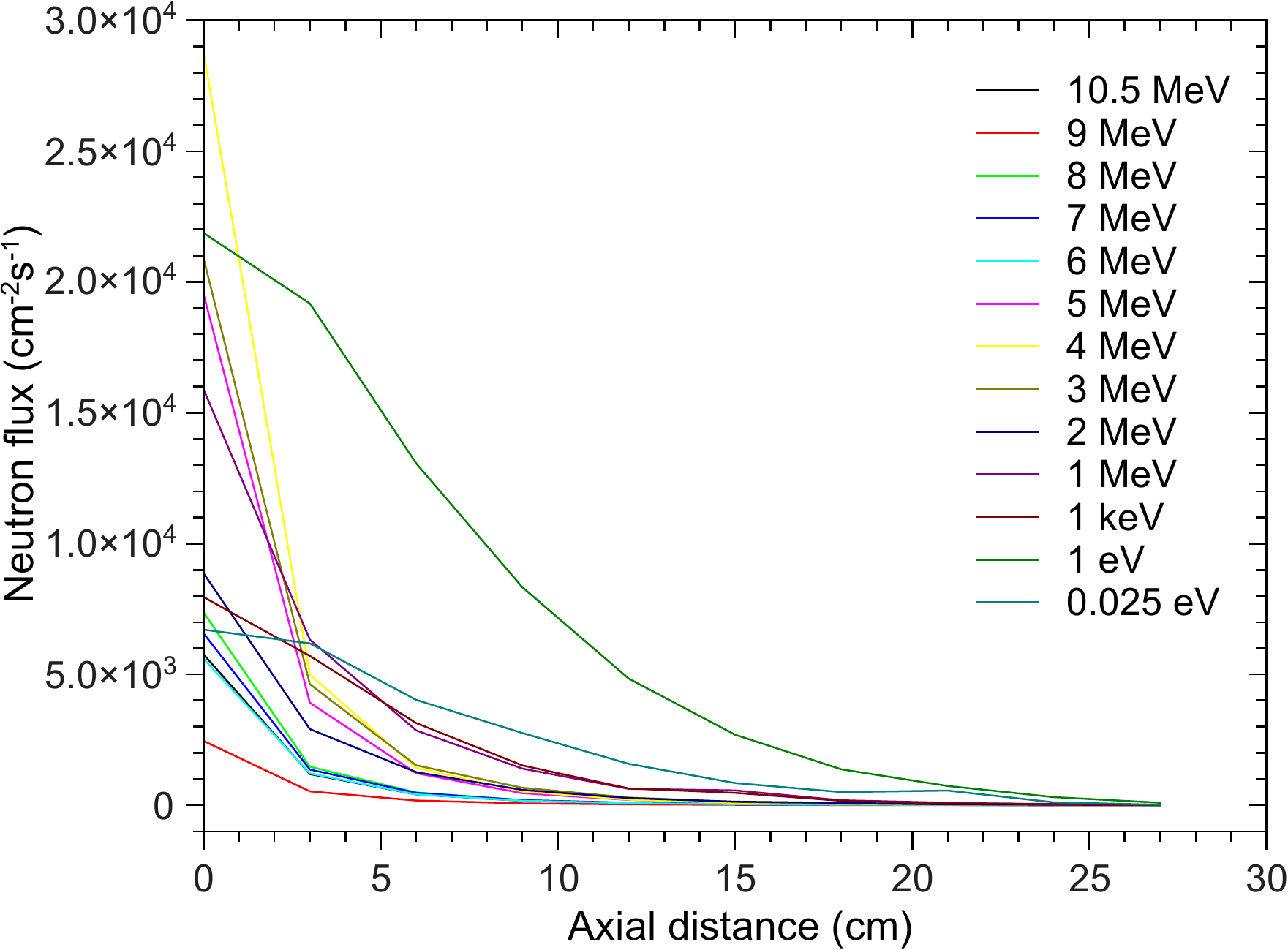}
\caption{Axial variation of the neutron flux for the defined energy bins.}
\label{energyvar}
\end{center}
\end{figure}
The latter plot for the same list of the energy bins is dedicated to the axial variation of the neutron flux for each energy bin as depicted in Fig.~\ref{energyvar}. Among the defined energy bins, the neutron flux of the 4-MeV energy bin attains the highest value, which is followed by that of the 1-eV energy bin, at the initial position on the plexiglass bar, reminding that the mean energy of neutrons emitted by the 5-Ci ${}^{239}$Pu-Be neutron source is 4.24 MeV.  From Fig.~\ref{energyvar}, we furthermore see that the lowest decrease rates of the neutron flux occur at the energy bins of 0.025 eV and 1 eV, which also means that the thermal energy interval sustains till a greater depth in comparison with the fast and epithermal energy intervals. To better support this statement, we plot the stacked bar chart of the specified energy bins in Fig.~\ref{stack}. From the bar segments shown in Fig.~\ref{stack}, it is seen that the energy bins that appertain to the fast energy interval constitute the largest group at the first spatial point; in contrast, the components of the fast neutron regime rapidly decay and become almost insignificant especially after a depth of 10 cm. When the behaviour of the thermal group is examined, we clearly show that the elements of the thermal energy interval remain evident within a distance of 18 cm and progressively become the majority during the propagation towards the outlet of the horizontal irradiation channel.
In the next step, we analyse the axial variation of the neutron flux in the plexiglass bar by using the conventional coarser group structure that is commonly defined as a set of the thermal, epithermal, and fast energy groups. The axial variation of the thermal group is illustrated in Fig.~\ref{thermepifastot} (a), and we reveal that the magnitude of the thermal flux is maintained in the order of $10^{4}$ till  a distance of 10 cm. Upon our MCS, the maximum thermal flux is determined as $2.86\pm0.04\times10^{4}$ $\rm cm^{-2}s^{-1}$, whereas this value is stated as $3\times10^{4}$ $\rm cm^{-2}s^{-1}$ in one of the initial studies with the NH3 howitzer~\cite{acti1}. Among the three energy groups, the neutron flux values of the epithermal group are always inferior compared to the thermal and fast energy groups as shown in Fig.~\ref{thermepifastot} (b), but the epithermal group furthermore brings computational challenge as can be seen from the uncertainties. Fig.~\ref{thermepifastot} (c) describes the spatial change of the fast neutron flux, and we demonstrate that the fast neutron flux decreases by one order of magnitude within a distance of 5 cm, whereas the same term for the thermal energy group reduces by one order of magnitude within a distance of 18 cm. Finally, in Fig.~\ref{thermepifastot} (d), the total neutron flux varying along the axial distance of the plexiglass bar in the NH3 howitzer is represented, and it is shown that the cumulative population of neutrons exponentially decreases onward the plexiglass bar. From our MCS results, we observe that the total flux of the neutrons entering the horizontal irradiation channel is $1.59\pm0.01\times10^{5}$ $\rm cm^{-2}s^{-1}$, while that of the neutrons leaking out the plexiglass bar is $2\pm0.25\times10^{2}$ $\rm cm^{-2}s^{-1}$. This drastic reduction in the neutron population along the axial distance in the plexiglass bar relatively demonstrates the neutron thermalisation and shielding capability of the plexiglass materials. 
To further investigate the variation of the neutron regimes, we track the ratio between the flux of the above-mentioned energy groups and the total neutron flux as shown in Fig.~\ref{ratio}. We can understand from Fig.~\ref{ratio} that, although the fast neutron cluster prominently predominates in the first few cm of the plexiglass bar, the fast neutron population fades through the neutron interactions such as down-scattering. Reminding that the composition of the plexiglass bar consists of 8 wt.$\%$H, 32 wt.$\%$O, and 60 wt.$\%$C, the presence of these low-A elements enhances the neutron moderation via the neutron elastic scattering, thereby enabling the transition from the fast group to the thermal group; however, a certain number of neutrons also leak out the plexiglass bar due to the scattering process. In the meanwhile, as illustrated in Fig.~\ref{ratio}, we see that the population of the thermal neutrons becomes the largest member of the overall neutron cluster after a depth of approximately 5 cm. The constituents of the plexiglass bar, especially hydrogen and carbon, furthermore have a substantial capability for the thermal neutron absorption. Thus, the thermal energy group loses a portion of its population due the thermal neutron absorption process. Finally, the ratio between the epithermal neutron flux and the total neutron flux is depicted in Fig.~\ref{ratio}, and we remark that epithermal population exhibits minor oscillations and always constitutes the smallest portion throughout the plexiglass bar. 
Lastly, as seen in Fig.~\ref{coluncol} and  Fig.~\ref{coluncolratio}, the final two plots for the neutronic analysis of the horizontal beam port are dedicated to the collided and uncollided neutron fluxes that are obtained through the utilisation of the point detectors. 
First, Fig.~\ref{coluncol} displays the axial variation of the collided and uncollided fluxes, and we see that both of these terms exponentially decrease when progressing in the plexiglass bar; however, it is worth mentioning that, based upon our MCS, the flux of the uncollided neutrons entering the horizontal irradiation channel is $4.718\pm0.024\times10^{4}$ $\rm cm^{-2}s^{-1}$, whereas the flux of the uncollided neutrons leaking out the plexiglass bar is $6.862\pm0.005$ $\rm cm^{-2}s^{-1}$. 
Secondly, Fig.~\ref{coluncolratio} shows the ratio between the collided neutron flux and the uncollided neutron flux. From the similarity between the collided/uncollided neutron flux ratio in Fig.~\ref{coluncolratio} and the thermal/total neutron flux ratio in Fig.~\ref{ratio}, we again understand that the thermal neutron population gradually becomes dominant by dint of neutron collisions within the plexiglass bar of the horizontal beam port.
\section{Conclusion}
In the current study, after a series of Monte Carlo simulations, we observe that the maximum thermal neutron flux as well as the maximum fast neutron flux occurs at the beginning of the plexiglass bar in the horizontal irradiation channel of the NH3 howitzer, and this observation implies the most practical region for the irradiation purpose since the irradiation time decreases when the neutron flux of the objective energy group increases. While the neutron flux of the thermal regime maintains its initial order of magnitude for a distance of 10 cm, the fast neutron flux drastically decays within few centimeters of the plexiglass bar. Finally, we also expose the capability of the plexiglass materials for the neutron flux reduction, and the present study shows that the plexiglass bar in the horizontal irradiation channel of the NH3 howitzer is able to lower the initial neutron flux by approximately three orders of magnitude.
\bibliographystyle{elsart-num}
\nocite{*}
\bibliography{Howitdosebiblio}
\end{document}